\documentclass[prl,twocolumn,superscriptaddress]{revtex4}
\usepackage{graphicx}
\usepackage[letterpaper, margin=1in]{geometry}
\usepackage{amsmath}
\usepackage{bbold}
\newcommand{\ket}[1]{\left|#1\right>}
\newcommand{\bra}[1]{\left< #1 \right|}

\newcommand{\ua}{\uparrow}
\newcommand{\da}{\downarrow}
\newcommand{\HH}{\hat{H}}
\newcommand{\beq}{\begin{equation}}
\newcommand{\eeq}{\end{equation}}
\newcommand{\bea}{\begin{eqnarray}}
\newcommand{\eea}{\end{eqnarray}}
\begin{document}
\title{Spin blockade as a probe of Zeeman interactions in hole quantum dots}
\author{Jo-Tzu Hung}
\affiliation{School of Physics, The University of New South Wales, Sydney NSW 2052, Australia}
\author{Elizabeth Marcellina}
\affiliation{School of Physics, The University of New South Wales, Sydney NSW 2052, Australia}
\author{Bin Wang}
\affiliation{School of Physics, The University of New South Wales, Sydney NSW 2052, Australia}
\affiliation{University of Science and Technology of China, Hefei, Anhui, 230026, China}
\author{Alexander R. Hamilton}
\affiliation{School of Physics, The University of New South Wales, Sydney NSW 2052, Australia}
\author{Dimitrie Culcer}
\affiliation{School of Physics, The University of New South Wales, Sydney NSW 2052, Australia}
\date{\today}
\begin{abstract}
Spin-orbit coupling is key to all-electrical control of quantum-dot spin qubits, and is frequently stronger for holes than for electrons. Here we investigate Pauli spin blockade for two heavy holes in a gated double quantum dot in an in-plane magnetic field. The interplay of the complex Zeeman and spin-orbit couplings causes a blockade leakage current anisotropic in the field direction. The period of the anisotropic leakage is critically dependent on the relative magnitude of Zeeman interaction terms linear and cubic in the magnetic field. The current and singlet-triplet exchange splitting can be effectively adjusted by an appropriate choice of field direction, providing a simple control variable for quantum information processing and a way of tailoring magnetic interactions in hole spin qubits.
\end{abstract}
\maketitle

Spin-based quantum information processing platforms relying on hole quantum dots (QDs) have recently attracted considerable attention \cite{Brunner_Science09,Zwanenburg_NanoLett09, DeGreve_NatPhys11,Greilich_NatPhoton11,Spruijtenburg_APL13, Li_APL13, Pribiag_NatNano13, Li_NanoLett15, Mueller_APL15,Voisin_NanoLett16, Brauns_PRB16b}, since they permit long spin coherence and electrically driven spin resonance thanks to the strong hole spin-orbit (SO) interaction \cite{Bulaev_PRL05, Bulaev_PRL07, Fischer_PRB08, Trif_PRL09, Fischer_PRL10, Wang_PRL12, Szumniak_PRB13, Maier_PRB13, Climente_NJP13}. Owing to their effective spin $J\!=\!\frac{3}{2}$, spin dynamics in hole systems often exhibits physics not found in electron systems \cite{Luttinger_PR55,Luttinger_PR56, Winkler_SST08,Culcer_PRL06, Culcer_PRL07, Kernreiter_PRB13}. Experimental progress in realizing high-quality two-dimensional (2D) and even lower-dimensional hole systems has opened the door to hole-based computing architectures \cite{Kloeffel_AR13, Salfi_Nanotech16,Salfi_PRL16,Huang_preprint16}. Probing the strengths of SO coupling and hole-hole interactions is thus highly relevant for quantum computing. Pauli spin blockade \cite{Hanson_RMP07, Ono_Science02}, the blocking of charge transport through QDs due to the Pauli exclusion principle, may be employed to perform this task in InSb, Si, Ge-Si core-shell wires, and GaAs \cite{Pribiag_NatNano13, Li_NanoLett15, Brauns_PRB16b, Wang_preprint16}.

Pauli spin blockade (PSB) is lifted by spin-flip processes most commonly originating in the SO or hyperfine interactions. Electron PSBs at low magnetic fields are primarily lifted by the hyperfine coupling to the nuclei \cite{Ono_PRL04, Koppens_Science05, Koppens_Nature06}, unless the singlet-triplet splitting is large due to a sizable interdot tunnel coupling \cite{Danon_PRB09, Nadj-Perge_PRB10,Nadj-Perge_PRL12}. Electron SO interaction only becomes a major lifting mechanism at stronger magnetic fields and strong interdot tunneling \cite{Nadj-Perge_PRB10}. For hole QDs, in contrast, the strong SO interaction is expected to be the dominant blockade lifting mechanism \cite{Pribiag_NatNano13, Li_NanoLett15, Brauns_PRB16b, Wang_preprint16} even at low magnetic fields, particularly due to the suppression of hole contact hyperfine interaction. Spin-flip cotunneling processes may also give rise to a leakage current \cite{Brauns_PRB16b}.

Here we investigate PSB in a gate-defined hole double QD in an in-plane magnetic field. We derive an effective SO tunneling Hamiltonian between $(1,1)$ and $(0,2)$, with $(N_L,N_R)$ charge state on the left and right dot. Our work shows that the PSB is anisotropic in the field orientation and strongly influenced by the complex Zeeman interaction, which in hole QDs may have terms both linear and cubic in the field strength. We find that the period of the anisotropic leakage is determined by the dominant terms in the Zeeman interaction. Based on this finding, one will be able to determine the form and magnitude of the Zeeman coupling of hole QDs with in-plane magnetic fields. The phonon induced relaxation assisted by the SO interaction may contribute to the leakage with a rate depending on the field orientation as well.

{\it Theoretical framework.} ---The effective spin $J\!=\!\frac{3}{2}$ comprises a heavy hole (HH) with a secondary quantum number $m_J\!=\!\pm\frac{3}{2}$ and a light hole (LH) with $m_{J}\!=\!\pm\frac{1}{2}$. For 2D holes, the HH-LH degeneracy at the band edge $\mathbf{k}\!=\!0$ originating in the bulk is lifted by the confinement in the growth direction $\parallel \hat{\bm z}$. Theoretical treatments often take this HH-LH splitting to be the largest energy scale ($>$ 10 meV in GaAs inversion layers), whereupon the lowest energy (HH) subband may be described by a pseudospin degree of freedom. While such perturbative treatments may break down in certain parameter regimes \cite{Marcellina_preprint16a}, they enable direct comparisons between low-dimensional hole and electron systems, and we adopt this picture here. Accordingly, the pseudospin Zeeman interaction has a complex form \cite{Winkler}. In the four-fold subspace describing the lowest energy subband the in-plane $g$-factor of the $m_{J}\!=\!\pm\frac{3}{2}$ states vanishes, and the leading-order Zeeman coupling is cubic in $B$ \cite{Winkler_PRL00}. Coupling to higher orbitals leads to a spin splitting linear in $B$ \cite{Winkler,Bulaev_PRL07}. In principle the coupling constants for both the $B$-linear and $B$-cubic interactions need to be determined for individual structures, and the 2D in-plane $g$ factor is typically small or nearly zero \cite{Rahimi_PRB03,Yuan_APL09,Marcellina_preprint16b}. SO contributions are important in lower dimensions. The Rashba interaction appears when the confinement potential is inversion asymmetric in the $\hat{\bm z}$ direction, and is tunable by the gate field \cite{Silov_APL04, Spruijtenburg_APL13, Wang_PRB13, Nichele_PRB14}. Dresselhaus SO terms due to bulk and interface inversion asymmetry are strong in 2D hole gases \cite{Durnev_PRB14}. All of these are strongly affected by the HH-LH splitting \cite{Wenk_PRB16}.

We consider two HHs confined in a 2D gated double dot grown along $\hat{z} \parallel [001]$ and positioned at $(x_j, y_j)\!=\!(\mp d,0)$, where $j\!=\!L,R$ label the left and right dot, respectively. In an in-plane magnetic field $\mathbf{B}\!=\!B(\cos{\theta},\sin{\theta})$ with $\theta$ measured from the $\hat{x}$ axis, the Hamiltonian is $\HH=\sum_{j=L,R}\HH^{(j)}_{\text{d}} +\HH^{(j)}_{\text{Z}}+\HH^{(j)}_{\text{SO}}$. Here, $\HH^{(j)}_{\text{d}}\!=\!\frac{(\mathbf{p}_{j}- e\mathbf{A}_{j})^2}{2m^{*}}+\frac{m^{*}}{2}[\omega^2_x (x-x_j)^2+\omega^2_y y^2]$ contains the kinetic energy and parabolic confinement, with $m^{*}$ the effective HH mass, $\mathbf{p}_j$ the canonical momentum, $\mathbf{A}_j$ the gauge potential, and $\hbar\omega_{x/y}$ the confinement energy quantum. The $j$ dot orbitals are the wavefunctions of the two harmonic oscillators $\ket{n_x,n_y^{(j)}}$ along the $\hat{x}$ and $\hat{y}$ axes, with $n_{x}$ and $n_y$ the respective quantum numbers. The dot shape is defined by two radii $R_{0,x/y}\equiv[\hbar/(m^{*}\omega_{x/y})]^{1/2}$, and for a circular QD, $R_{0,x}=R_{0,y}=R_{0}$. 

For the Zeeman Hamiltonian $\HH^{(j)}_{\text{Z}}$ we include both $B$-linear and $B$-cubic terms \cite{Winkler}, so that
\bea
\HH^{(j)}_{\text{Z}} &=& -\frac{3}{2}q\mu_{B}B\left( e^{i\theta}\sigma_{j+}+e^{-i\theta}\sigma_{j-}\right)\\
&&-\, \mu_B B \frac{f}{\hbar^2}\left(e^{-i\theta}\sigma_{j+} p^{2}_{j-}+e^{i\theta}\sigma_{j-} p^{2}_{j+}\right)\nonumber\\
&&+\, F(\mu_B B)^3 \left( e^{3i\theta}\sigma_{j+}+e^{-3i\theta}\sigma_{j-}\right)\nonumber\,,\label{eq:zeeman}
\eea
with $\boldsymbol{\sigma}_{j}$ the Pauli matrices of the $j$ pseudospin, $\sigma_{j\pm}\!\equiv\!\frac{1}{2}(\sigma_{jx}\pm i\sigma_{jy})$ and $p_{j\pm}\!\equiv\!-i\hbar (\partial_{jx}\pm i\partial_{jy})$. The first term in $\HH^{(j)}_{\text{Z}}$ is the direct linear coupling, the second term an indirect linear coupling allowed by the symmetry \cite{Pryor_PRL06}, while the third represents the indirect cubic term \cite{Winkler}. The material-dependent parameters $f$ and $F$ are both inversely proportional to the HH-LH splitting \cite{Supp_fF}. 

\begin{figure}[t]
	\includegraphics[width=0.9\linewidth]{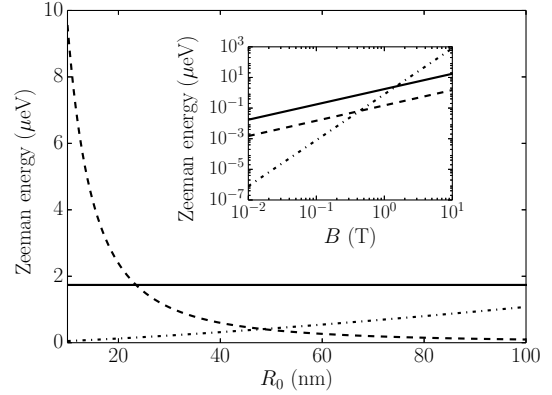}
\caption{Contributions to the Zeeman energy of a GaAs double QD at $B\!=\!1$ T as a function of the dot radius $R_0$, due to the direct linear coupling (solid line), the indirect $B$-linear term (dashed line) and $B$-cubic coupling (dot-dashed line.) We have assumed HH-LH splitting $10$ meV and included only the first $\hat{z}$ subband. Inset: individual contributions for $R_0=80$ nm, as a function of $B$.}\label{fig:ss}
\end{figure}

Figure~\ref{fig:ss} shows the three individual contributions to the Zeeman energy of a double dot. While the direct $B$-linear contribution is independent of the dot size, the indirect $B$-linear ($B$-cubic) contribution will be reduced (enhanced) when $R_0$ grows. The $B$-cubic contribution can be dominating at large $B$.

We adopt the single-spin basis comprised by $\ket{\ua}_{\theta_Z}\!=\!\frac{1}{\sqrt{2}}[1,e^{-i\theta_{Z}}]^{T}$ and $\ket{\da}_{\theta_Z}\!=\!\frac{1}{\sqrt{2}}[1,-e^{-i\theta_{Z}}]^{T}$, with $\theta_{Z}$ subject to the dominating Zeeman term and $[\cdots]^T$ the transpose of a row vector. If the direct $B$-linear term is the main contribution to the Zeeman splitting, $\theta_{Z}\!=\!\theta$. Otherwise, $\theta_{Z}\!=\! -\theta$ ($-3\theta$) in the case of the dominating indirect $B$-linear ($B$-cubic) term. We expand the two-spin Hilbert space into the singlet $\ket{S}\!=\!\frac{1}{\sqrt{2}}(\ket{\ua\da}_{\theta}-\ket{\da\ua}_{\theta})$, and the unpolarized and two polarized triplet states, $\ket{T_0}\!=\!\frac{1}{\sqrt{2}}(\ket{\ua\da}_{\theta}+\ket{\da\ua}_{\theta})$, $\ket{T_{+}}\!=\!\ket{\ua\ua}_{\theta}$ and $\ket{T_{-}}\!=\!\ket{\da\da}_{\theta}$.

The SO interaction $\HH^{(j)}_{\text{SO}}=\HH^{(j)}_{\text{R}}+\HH^{(j)}_{\text{D}}$, with
\bea
\!\!\!\!\!\!\HH^{(j)}_{\text{R}}&=& i\alpha(\sigma_{j+} p^{3}_{j-} - \sigma_{j-} p^{3}_{j+})\,,\label{eq:R}\\
\!\!\!\!\!\!\HH^{(j)}_{\text{D}}&=&-\beta_{3}  (\sigma_{j+} p_{j-}p_{j+}p_{j-} + \sigma_{j-} p_{j+} p_{j-}p_{j+})\nonumber\\
	&&-\, \beta_{1}(\sigma_{j+} p_{j-}+ \sigma_{j-} p_{j+}) \label{eq:D} \,.\quad 
\eea
Here $\alpha$ ($\beta_{1,3}$) is the Rashba (linear and cubic Dresselhaus) coupling strength.

\begin{table}[b]\centering
\caption{The values of $t_0$, and $t_R$ based on the results $\alpha$ of 2D accumulation (A) and inversion (I) holes \cite{Marcellina_preprint16a}. For 2D Si, there is a significant anisotropic Rashba contribution in $k_{\parallel}$, and for hole densities $\sim 10^{16}$ m$^{-2}$, the ratio of the Rashba spin splitting to the Fermi energy is roughly $0.1$, hence $t_{\text{R}}\sim t_0/10$.}
\label{tab:parameters}
\begin{tabular}{lcccc}
\hline\hline
					&GaAs	&InAs	&InSb	&Si	\\
\hline
	$R_0$ (nm)		&$30$	&$38$	&$50$	&$20$\\
\hline
	$t_{0}$ ($\mu$eV)	&230		&180		&97		&530	\\		
\hline
	$\alpha$ (meV$\cdot$m$^3$) &$5.4\times 10^{-25}$ & $3.6\times 10^{-24}$ & $1.0\times 10^{-23}$ & -\\
\hline	
	$t_{\text{R}}$ ($\mu$eV)	&150(A)/50(I)	&390	(I) 	&603(I)	& -\\						
\hline\hline		
\end{tabular}
\end{table}
%
{\it Spin-orbit tunneling direction.} ---Tunneling may arise from spin-preserving and spin-flip processes. The former corresponds to the conventional interdot tunneling strength $t^{nm}_0\!\equiv\!\bra{^{L}n_x,n_y}\HH^{(L)}_{\text{d}}\ket{m_x,m_y^R}$. Similarly, we have for the SO interaction $\bra{^{L}n_x,n_y}\HH^{(L)}_{\text{SO}}\ket{m_x,m_y^R}$, causing a preferential direction in the spin Hilbert space, which we shall refer to as the tunneling field direction \cite{Takahashi_PRL10,Nadj-Perge_PRL12}. If $\mathbf{B}$ points along the tunneling field direction, there is no SO tunneling. For the double dot aligned along $\hat{x}$, if $n_y\!=\!m_y$, the Rashba tunneling field is along $\hat{y}$, and linear and cubic Dresselhaus tunneling fields both point along $\hat{x}$. 

{\it Effective Hamiltonian.} ---The usual PSB involves the (0,2) singlet $\ket{S_{02}}$, and the $(1,1)$ states $\{\ket{S},\ket{T_0},\ket{T_{\pm}}\}$. As $\mathbf{B}$ rotated, the tunneling matrix elements $\Delta_0\!=\!\sqrt{2}(it_{R}\sin{\theta}\!+\!it_{D}\cos{\theta}\!-\!t_B\cos{2\theta})$ between $\ket{S_{02}}$ and $\ket{T_0}$, and $\Delta_{\pm}\!=\!(t_{R}\cos{\theta}\!-\!t_{D}\sin{\theta}\!-\!i t_B \sin{2\theta})$ between $\ket{S_{02}}$ and $\ket{T_{\pm}}$ will vary in strength. Here we have used $t_R\!\equiv\!\bra{^L0,0}\HH_{\text{R}}\ket{0,0^R}$, $t_D\!\equiv\!\bra{^L0,0}\HH_{\text{D}}\ket{0,0^R}$, $t_B\!=\!f\mu_BB\chi/R^2_0$, and the orbital overlap $\chi\!\equiv\!\langle^{L}0,0|0,0^R\rangle\!=\!e^{-d^2/R^2_0}$. Because $\ket{S}$ and $\ket{T_0}$ are degenerate, we adopt their superpositions $\ket{M}\!\equiv\!\frac{1}{N_{M}}\left[\Delta^{*}_{0}\ket{S}\!-\!t_{0}\ket{T_{0}}\right]$ and $\ket{M_{\perp}}\!\equiv\!\frac{1}{N_{M}}(t_{0}\ket{S} +\Delta_0\ket{T_{0}})$, with $t_0\equiv t^{00}_0$ and $N_{M}\!=\!\sqrt{|\Delta_{0}|^{2}+t^{2}_{0}}$. 

In the $\{S_{02},M,M_{\perp},T_{+},T_{-}\}$ basis, the effective tunneling Hamiltonian is
\beq
	\HH_{\text{eff}} = \left(\begin{array}{ccccc}
	-\epsilon		& 0	& N_{M} & \Delta_{+} & \Delta_{-}  \\
	0			& 0	& 0	& 0  & 0 \\
	N_{M}			& 0	& 0	& 0 & 0   \\
	\Delta^{*}_{+}	&0	&0	& E_{Z} & 0  \\ 
	\Delta^{*}_{-} 	&0	& 0	&0   & -E_{Z}
	\end{array}\right)\,,\label{eq:Heff}
\eeq
with $\epsilon$ the detuning between the charge states $(0,2)$ and $(1,1)$, and $E_{Z}\!=\! -\mu_BB[3q+\frac{2f\chi^2}{R^2_0}\cos{(2\theta)}]$ the Zeeman splitting. From Eq.~(\ref{eq:Heff}), if the two HHs are initialized in $\ket{M}$, there will be no leakage without spin relaxation among the $(1,1)$ states. Experimentally, the initial state is often unknown, and there may exit the $(1,1)$ relaxation process.
\begin{figure}[t]
	\includegraphics[width=0.9\linewidth]{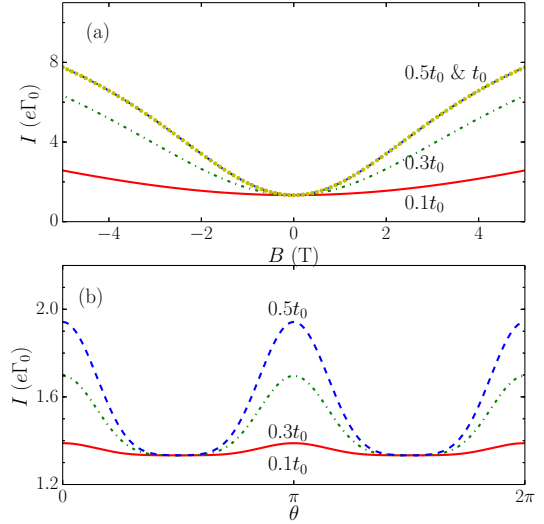}
\caption{GaAs double QD with $R_0=30$ nm and $t_0\!=\!200$ $\mu$eV. (a) $I(B)$ at $\theta=0$, for various values of $t_R/t_0$. The $t_R\!=\!0.5$ and $t_R\!=\!t_0$ curves are overlapping. (b) $I(\theta)$ at $B=1$ T. $\Gamma_{11}\!=\!\Gamma_0=3$ MHz, making $e\Gamma_0\!\approx\!0.48$ pA.}\label{fig:psb1}
\end{figure}
\begin{figure}[t]
	\includegraphics[width=0.95\linewidth]{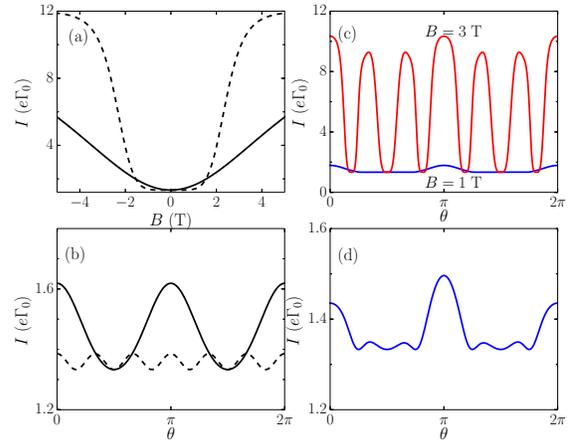}
\caption{GaAs double QD with $R_0=80$ nm, $t_R\!=\!0.5t_0$ and $\Gamma_{11}\!=\!\Gamma_0=3$ MHz. (a) $I(B)$ with $\theta\!=\!0$ due to the $B$-cubic term (dot-dashed line) and due to the $B$-linear terms (solid line). (b) $I(\theta)$ at $B=1$ T due to the $B$-cubic term (dot-dashed line) and due to the $B$-linear terms (solid line). (c) The beat pattern in $I(\theta)$, when both the $B$-cubic and $B$-linear terms included. (d) $I(\theta)$ at $B\!=\!1$ T, when the spin splitting in the left and right dot is linear and cubic in $B$, respectively.}\label{fig:psb2}
\end{figure}

Table~\ref{tab:parameters} lists the calculated values of $t_0$ and $t_R$ for typical semiconductors. $R_0$ is chosen so that the HH-LH mixture is a perturbation. We have assumed a negligible interdot distance, i.e., $\chi\approx e^{-1}$, but note that in GaAs, $t_{\text{R}}/t_0$ for holes may be much higher than for electrons \cite{Nadj-Perge_PRB10}.
%

{\it Spin-orbit induced leakage.} --- The leakage is determined by a final readout of the tunneling from $(0,2)\!\rightarrow\!(0,1)$ \cite{Supp_SOL}. Charge flow from $(1,1)\!\rightarrow\!(0,2)\!\rightarrow\!(0,1)$ is balanced by the transition rates between the seven states, including the $(1,1)$ relaxation rate $\Gamma_{11}$. We obtain the eigenstates $\{\ket{k}\}$ of Eq.~(\ref{eq:Heff}), and calculate the leakage by solving a set of steady-state kinetic equations $d\mathcal{P}/dt\!=\! -\mathcal{W}_{\text{out}}\mathcal{P}+\mathcal{W}_{\text{in}}\mathcal{P}\!=\!0$, where $\mathcal{P}\!=\![\{P_{k}\},P_{\ua},P_{\da}]^{T}$ is a vector consisting of the probabilities $P_k$ in the eigenstates $\{\ket{k}\}$ and $P_{\sigma}$ in $(0,1)$ with the spin $\sigma$. The transition rates into and out of each state are contracted in the respective matrices $\mathcal{W}_{\text{in}}$ and $\mathcal{W}_{\text{out}}$. We numerically solve $\mathcal{P}$ and obtain the leakage current $I\!=\!e\Gamma_{DL}P_{02}$, with $\Gamma_{DL}$ the dot-lead transition rate and $P_{02}$ the final probability to end in $(0,2)$. The following results are shown for $\epsilon\!=\!0$ and $t_R\!\gg\!t_D$. $\Gamma_{11}=\Gamma_0$ is first set constant to focus on the SO-induced leakage.

Figure~\ref{fig:psb1} shows the leakage $I(B)$ and $I(\theta)$ in GaAs, when the $B$-linear Zeeman couplings included only. The PSB is lifted at $B\!\neq\!0$ by the tunable Rashba SO coupling, and $|dI/dB|$ grows when $t_R$ increased. While we have solved the leakage $I$ numerically, if $t_R,E_{Z}< t_0$ and $\Gamma_{11}$ is $B$-independent, $I(E_Z)\!\sim\!e\Gamma_{\text{DL}}P_{M}\gamma^2[1-\Lambda^2/(E^{2}_{Z}+\Lambda^2)]$, with $\Lambda\!\equiv\!\gamma t^2_0/|\Delta_{+}(\theta)|$, $\gamma^2\!=\!\Gamma_{11}/\Gamma_{DL}$ and $P_{M}$ the probability in $\ket{M}$. This approximation indicates $I(B)$ is a Lorentzian \cite{Danon_PRB09}. In Fig.~\ref{fig:psb1}(b), the maximal (minimal) $I(\theta)$ is found when $|\Delta_{+}|\!\approx\!\ (t^2_{R}\cos^2{\theta}+t^2_B\sin^2{2\theta})^{1/2}$ reaches its maximum (minimum). In Fig.~\ref{fig:psb2}(a)-(b), when only the $B$-cubic term included, the lifting is insignificant at low $B$, and the period of $I(\theta)$ is $1/3$ of that in the $B$-linear case. In Fig.~\ref{fig:psb2}(c), we consider both the $B$-linear and $B$-cubic couplings, and find a beat pattern in $I(\theta)$ at large $B$ due to the two competing Zeeman terms. Figure~\ref{fig:psb2}(d) is the beat pattern of having the $B$-linear splitting in the left dot and the $B$-cubic splitting in the right dot, and the left dot has a larger spin splitting. 

If the $(1,1)$ states are well split by $B$, $\Gamma_{11}$ is spin-selective, and the number of the relaxation channels is reduced. For III-V QDs with a large HH-LH splitting, the major cause of spin relaxation is the SO assisted phonon relaxation \cite{Bulaev_PRL05,Trif_PRL09,Climente_NJP13}. Indeed, the SO effect may vary in strength when $\mathbf{B}$ is rotated, however, when $\HH^{(j)}_{\text{SO}}$ yields multiple sources of SO tunneling, spin mixing is present at all $\mathbf{B}$. There are two-spin and single-spin relaxation channels. For QDs with small SO mixture, including two-spin relaxation is sufficient because the current dot orbital is well decoupled from the others. Such two-spin relaxation requires nonvanishing $\Delta_{\pm}(\theta)$. If the SO mixture is large, corrections from single-spin relaxation should be included. 

\begin{figure}[t]
	\includegraphics[width=0.95\linewidth]{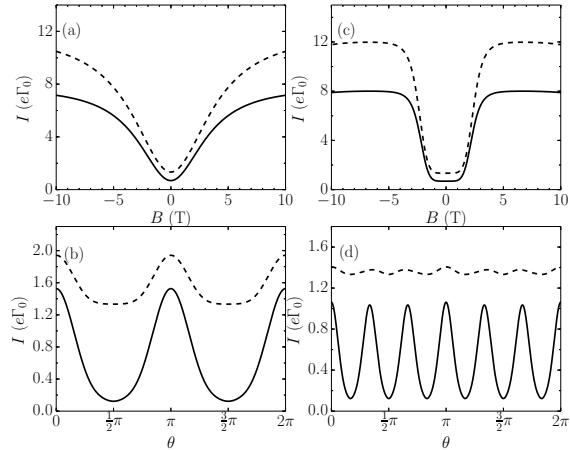}
\caption{The leakage $I(B,\theta)$ with $t_R=0.5t_0$ with spin-selective $\Gamma_{11}$ (solid line) in GaAs. The dashed lines are obtained with a constant $\Gamma_{11}\!=\!\Gamma_0=3$ MHz. (a)-(b) At $R_0=30$ nm, $I(B)$ with $\theta=0$, and $I(\theta)$ at $B=1$ T, due to the $B$-linear coupling. (c)-(d) At $R_0\!=\!80$ nm, $I(B)$ with $\theta\!=\!0$ and $I(\theta)$ at $B\!=\!1$ T, due to the $B$-cubic term.}\label{fig:gammaB}
\end{figure}

We adopt the $S$-$T_{\pm}$ relaxation rate $\Gamma^{ST_{\pm}}$ and single-spin one $\Gamma^{\Uparrow\Downarrow}$. To compare with the results with $\Gamma_{11}\!=\!\Gamma_0$, we set $\Gamma^{ST_{\pm}}\approx\Gamma_{0}|\cos{\theta}|^2$ ($\Gamma_{0}|\cos{3\theta}|^2$ in the $B$-cubic case). Assuming the single-spin relaxation assisted by the Dresselhaus interaction, we set $\Gamma^{\Uparrow\Downarrow}(B)\!\propto\!B^2$ \cite{Supp_Gamma} with $\Gamma^{\Uparrow\Downarrow}(B)=\Gamma_{0}$ at $\mathbf{B}\!=\!1$ T $\hat{y}$ and vanishes when $\mathbf{B}\parallel \hat{x}$. In Fig.~\ref{fig:gammaB}(a) and (c), $I(B)$ has a smaller profile, due to the reduced number of the relaxation channels, whereas $I(\theta)$ in Fig.~\ref{fig:gammaB}(b) and (d) have a larger amplitude, from the anisotropic $\Gamma_{11}$.
%

{\it Experimental applicability.} ---We have assumed $t_R\gg t_D$, however, if $t_D\!\sim\!t_R$, their relative strength may be determined by the minimum positions of $I(\theta)$. We have focused on the one-hole limit, whereas the existing HH PSB experiments are often performed with higher charge filling \cite{Li_NanoLett15, Wang_preprint16}. If multiple holes are occupied, the SO tunneling direction is sensitive to the dot shape and the symmetry of involved dot orbitals \cite{Wang_preprint16}. The other contribution to the leakage $I$ (albeit implicitly) is introduced by $\Gamma_{11}$. We have considered $\Gamma_{11}$ on the order of MHz, making the calculated $I$ on the order of pA \cite{Brauns_PRB16b,Wang_preprint16}. 

The anisotropic $I(\theta)$ can be manipulated by varying $\theta$ and employed in probing the power of $B$ in the Zeeman interaction via the signal period. Because $\Delta_{\pm}$ may be determined from the singlet-triplet anticrossing (along with $t_0$ extracted from the anticrossing at $\epsilon=0$), the strength $|E_{Z}|$ can be extracted via $I(B)$. A desired $I(\theta)$ could also be engineered for further computing applications. 

Moreover, $\ket{M}$ and $\ket{M_{\perp}}$ correspond to the two pseudospin $S_z\!=\!0$ states of the $(1,1)$ regime in the singlet-triplet qubit architecture \cite{Petta_Science05}. For hole double dots with a non-negligible $t_R/t_0$, the exchange splitting may be effectively tuned via $\theta$. 

Lastly, Ref.~\onlinecite{Marcellina_preprint16a} indicates that the perturbative approach will break down if the 2D hole density reaches a critical value. The current model then requires a lower bound for the dot radius $R_0$. The 2D results of GaAs accumulation (inversion) layers therein suggest $R_0\gtrsim 13 $ nm ($\gtrsim 5.6$ nm).

{\it Conclusions.} ---We have investigated the SO-induced PSB leakage in a two-HH double QD in an in-plane magnetic field. We have shown that the anisotropic leakage signal may be controlled by varying the magnetic field orientation and employed as a probe of the Zeeman coupling in hole QDs. We will extend our work with the full $J\!=\!\frac{3}{2}$ Hamiltonian to acquire more complete behavior of confined holes. 

\section*{Acknowledgements}
We thank Daisy Q. Wang for providing experimental data and useful discussion, Tetsuo Kodera and Andr{\'a}s P{\'a}lyi for fruitful discussions. This work is supported by the ARC through the DP scheme.

\appendix
\section{Appendix}
\subsection{THE SO COUPLING FORM}\label{app:so} 

Here we give the Rashba and cubic Dresselhaus couplings for the two HHs occupying on the ground orbital, $\ket{0,0^{(j)}}=(\pi R^2_0)^{-1}e^{-[(x-x_j)^2+y^2]/2R^2_0}$. The spin basis is given the bulk-$q$ Zeeman term.
\bea
\!\!\!\!\!\!\!\bra{S_{02}}\sum_{j}\HH^{(j)}_{R}\ket{T_{\pm}} &\!=\!& \pm \frac{\alpha \hbar^3}{R_{0}^3} \cos{\theta}\,e^{-\zeta ^2}\zeta^3\, \\
\!\!\!\!\!\!\!\bra{S_{02}}\sum_{j}\HH^{(j)}_{R}\ket{T_{0}} &\!=\!& i\frac{\sqrt{2}\alpha \hbar^3}{R_{0}^3} \sin{\theta}\,e^{-\zeta^2}\zeta^3\,,\\
\!\!\!\!\!\!\!\bra{S_{02}}\sum_{j}\HH^{(j)}_{D}\ket{T_{\pm}} &\!=\!&\mp \frac{\beta_{3} \hbar^3}{R_{0}^3}
		\sin{\theta}\, e^{-\zeta^2}\xi (2-\zeta^2),\quad\\
\!\!\!\!\!\!\!\bra{S_{02}}\sum_{j}\HH^{(j)}_{D}\ket{T_{0}} &\!=\!& i\frac{\sqrt{2} \beta_{3} \hbar^3}{R_{0}^3}
			\cos{\theta}\, e^{-\zeta^2}\zeta (2-\zeta^2),\quad\quad
\eea
where $\zeta \equiv d/R_0$ and $\chi = e^{-\zeta^2}$ is the orbital overlap.

\section{THE INDIRECT MAGNETIC COUPLINGS}\label{app:722}
According to Ref.~\onlinecite{Winkler}, the parameter $f=f_1+f_2$ from the indirect $B$-linear coupling has two contributions. The contribution $f_1$ concerns the HH-LH mixing at the same subband along the $\hat{z}$ direction, whereas $f_2$ involves the HH-LH mixing between different $\hat{z}$ subbands. The overall contribution $f$ is strongly affected by the confinement asymmetry in 2D hole systems.

Below we consider a quasi-triangular well and focus on the first subband $a=1$ only, i.e., $f_2 = 0$. We estimate $f_{1}$ using the Fang-Howard trial wavefunction $F_i(z)=2\lambda^{3/2}_{i}ze^{-\lambda_{i}z}$ with $\lambda_{i}=\lambda_{h}$ for HHs and $\lambda_{i}=\lambda_{l}$ for LHs \cite{Davies}. We obtain
\bea
	f_{1} &\approx& -\frac{3}{2}\kappa\bar{\gamma}\frac{\hbar^2}{m_0(E^h_1-E^l_1)}\frac{128\lambda^3_{h}\lambda^3_{l}}{(\lambda_h+\lambda_l)^6}\,.
\eea
with $m_0$ the free-electron mass and $E^{h (l)}_a$ the energy of the 2D HH (LH) state $\ket{h_{a}(l_{a})}$ at the subband $a$. We have assumed the axial approximation $\bar{\gamma}=\frac{1}{2}(\gamma_2+\gamma_3)$ with $\gamma_{1,2,3}$ the Luttinger parameters. For typical hole densities $\sim 10^{15}$ m$^{-2}$, the indirect $B$-linear effect corresponding to $f_1$ is on the order of $\mu$eV at $B=1$ T in GaAs.

The other indirect coupling cubic in $B$ is significant for large QDs and at high fields, and we refer to Eq.~(7.19) of Ref.~\onlinecite{Winkler} (and its errata) for the expression of $F$.

\subsection{THE STEADY-STATE KINETIC EQUATIONS}\label{app:master} 
The kinetic equations used in the main text are given by
\bea
\!\!\!\!\!\!\!&&\frac{dP_k}{dt}\!=\! \sum_{\sigma}\left(U_{\sigma k}\, P_{\sigma}-W_{k{\sigma}}P_{k}\right)\!+\!\sum_{k'\neq k}\left(\Gamma_{k'} P_{k'}-\Gamma_{k} P_{k}\right)=0 \,,\nonumber\\
\!\!\!\!\!\!\!&&\frac{dP_\sigma}{dt} \!=\! \sum_{k}\left(W_{k\sigma}P_{k}-U_{\sigma k}\,P_{\sigma}\right)=0\,,\label{eq:meq}
\eea 
where $\Gamma_{k}$ is the $(1,1)$ relaxation rate out of $\ket{k}$. $W_{k\sigma}$ and $U_{\sigma k}$, given as follows, correspond to the escape rates from $\ket{k}$ to $(0,1)_{\sigma}$ and the refilling rate from $(0,1)_{\sigma}$ to $\ket{k}$.
\bea
&&W_{k\sigma} = \Gamma_{R}\sum_{\sigma'} |\bra{(0,1)_{\sigma}} d_{R_{\sigma'}}\ket{k}|^{2}\,,\\
&& U_{\sigma k} = \Gamma_{L}\sum_{\sigma'} |\bra{k} d^{+}_{L_{\sigma'}}\ket{(0,1)_{\sigma}}|^{2}\,,	
\eea
with $\Gamma_{R(L)}$ the tunneling rate between the right (left) dot and lead, and $d_{j} (d^{+}_{j})$ the HH annihilation (creation) operator on the $j$ dot. 

\subsection{SIMPLIFIED EXPRESSIONS OF THE LEAKAGE CURRENT}\label{app:leakage} 

The leakage current is given by $I\!=\!e\Gamma_{R}P_{02}\!=\!e\sum_{\sigma,k}W_{k\sigma}P_{k}$.

When $B\!\rightarrow\!0$, only one state $\ket{u}$ is unblocked, making $W_{k\sigma}\!=\!0$ for $k\!\neq\!u$. We have $P(0,1)_{\sigma}\!=\!\sum_{k}(W_{k\sigma}P_k/U_{\sigma k})$ from Eq.~(\ref{eq:meq}), and write the probability in $\ket{u}$ by $P_{u}\!=\!\frac{1}{3\Gamma_u}\sum_{k\neq u}\Gamma_{k}P_{k}$. The resulting current becomes $I(B\rightarrow\!0)\!=\! \sum_{\sigma, k\neq u}\frac{eW_{u\sigma}}{3\Gamma_{u}}\Gamma_{k}P_{k}$. We note that $I(B\rightarrow0)$ is only nonzero due to a finite (1,1) relaxation rate $\Gamma_k$. For a rough estimate, we assume that $\ket{u}$ contains the amplitude $A_{u,02}$ of $\ket{S_{02}}$ so that $W_{u\sigma}\!=\! \frac{\Gamma_{R}}{2}|A_{u,02}|^2$, and obtain $I_0\sim e\Gamma_{\text{DL}} \frac{|A_{u,02}|^2}{3}\sum_{k\neq u}P_k$, with $\Gamma_{\text{DL}}\sim \Gamma_{L(R)}$ and $\Gamma_{u,k}\sim\Gamma_{11}$. 

For $B\!\neq\!0$, the probability in an unblocked state $\ket{u}$ is $P_{u}\!=\!P_{M}\sum_{\sigma}[\Gamma_{M}U_{\sigma u}(\Gamma_{u}U_{\sigma u}+ 2W_{\sigma M}W_{u\sigma})^{-1}]$. This leads to a leakage current
\bea
I&=& eP_{M}\sum_{u,\sigma}W_{u\sigma}\sum_{\sigma'}\frac{\Gamma_{M}U_{\sigma'u}}{\Gamma_{u}U_{\sigma'u}+2U_{\sigma M}W_{u\sigma'}}\,.\label{eq:I_B}\nonumber
\eea
For a {\it rough} estimate, we find
\beq
	I \sim e\Gamma_{\text{DL}} P_{M} \sum_{u}\frac{|A_{u,02}|^{2} |A_{u,11}|^{2} \gamma^2}{|A_{u,11}|^{2}\gamma^2+|A_{u,02}|^{2}}\,,\label{eq:I_approx}
\eeq
where $\gamma^2 \!\equiv\! \Gamma_{11}/\Gamma_{\text{DL}}$, and $U_{\sigma u}\!\equiv\! \frac{\Gamma_{L}}{2}|A_{u,11}|^{2}$ with $A_{u,11}$ the amplitude of the $(1,1)$ charge configuration in $\ket{u}$. At $\epsilon=0$, when the SO coupling is a perturbation to a large $t_0>E_{Z}$, we have $A_{T_{\pm},02}\propto  |\Delta_{\pm}|E_{Z}(E_Z^2-t^2_0)^{-1}$ and $A_{T_{\pm},11}$ barely depends on $E_{Z}$. As such, the leakage current $I \sim e\Gamma_{\text{DL}} P_{M}E^{2}_{Z}\gamma^2(E^{2}_{Z}+\frac{t^4_0}{|\Delta_+|^2}\gamma^2)^{-1}$ when $|E_{Z}|/t_0$ is small. This approximated $I(B)$ has a Lorentzian shape \cite{Danon_PRB09}.

\subsection{THE PHONON INDUCED RELAXATION}\label{app:rel} 
The hole-phonon interaction is \cite{Woods_PRB04,Bulaev_PRL05,Trif_PRL09,Climente_NJP13}
\beq
	\HH_{\text{ph}} =  \sum_{q\lambda} \frac{F(q_{z})e^{i\mathbf{q}_{\parallel}\cdot (\mathbf{r}_1+\mathbf{r}_2)}}{\sqrt{2\rho_c\omega_{q\lambda}/\hbar^2}} [P(\mathbf{q})-iD(\mathbf{q})] (b^{+}_{-\mathbf{q}\lambda}+b_{\mathbf{q}\lambda})\,,
\eeq
where $P(\mathbf{q})\!=\!e\beta_{\mathbf{q}\lambda}$ and $D(\mathbf{q})\!=\!(D_0\mathbf{q}\cdot{\boldsymbol{\xi}_{q\lambda}}-D_{z}q_{z}\xi^{z}_{q\lambda})$ result from the piezoelectric field $\beta_{\mathbf{q}\lambda}$ and deformation $D_{0/z}$. We have denoted by $\xi_{q\lambda}$ at the wave number $q$ and the vibration mode $\lambda$,  by $b^{+}_{\mathbf{q}\lambda}$ ($b_{\mathbf{q}\lambda}$) phonon creation (annihilation) operator, and by $\rho_c$ the crystal mass density. The factor $F(q_{z})$ is a unity within the dot height, otherwise it is vanishing.  

The hole-phonon interaction couples the double-dot orbitals through the phase factor $e^{i\mathbf{q}_{\parallel}\cdot (\mathbf{r}_1+\mathbf{r}_2)}$. For QD qubits, the dipole approximation $e^{i\mathbf{q}_{\parallel}\cdot (\mathbf{r}_1+\mathbf{r}_2)}\approx 1+ i\mathbf{q}_{\parallel}\cdot (\mathbf{r}_1+\mathbf{r}_2)$ is reasonable \cite{Huang_PRB14b} and yields analytic results. Within the approximation, the piezoelectric coupling ($\propto 1/\sqrt{q}$) is the dominant phonon relaxation mechanism.

For the singlet-triplet relaxation, the effective $S_{11}$ and $T_-$ are given by $\ket{S'_{11}}\!=\!a\ket{S_{11}} + c_{+}\ket{T_{+}} + c_{-}\ket{T_{-}}$ and $\ket{T'_{-}}\!=\!b\ket{S_{11}} + d_{-}\ket{T_{-}}$ with appropriate state coefficients $a$, $b$, $c_{\pm}$ and $d_{-}$. The effective $S_{11}$-$T_-$ relaxation rate due to the piezoelectric coupling is given by $\Gamma^{ST_-}=(|a^{*}b|^2+|c^{*}_{-}d_{-}|^2)\Gamma^{\text{PE,ST}}_{\lambda}$, where the base rate
\beq
	\Gamma^{\text{PE,ST}}_{\lambda} \approx \eta^{\text{PE}}_{\lambda}\frac{2(eh_{14})^2}{\hbar\rho_c\epsilon^2_{r}v^3_{\lambda}}\omega_{ST} [1+n(\omega_{ST})],\label{eq:rel_stm}
\eeq
with $eh_{14}=1.2\times 10^9$ V/m, $\epsilon_{r}$ the relative permittivity, $\hbar\omega_{ST}$ the singlet-triplet energy separation, and $v_{\lambda}=v_{l (t)}$ the longitudinal (transverse) acoustic speed. The dimensionless parameters $\{\eta^{\text{PE}}_{l},\eta^{\text{PE}}_{t1},\eta^{\text{PE}}_{t2}\}=\{\frac{12}{5},\frac{4}{21},\frac{1}{15}\}$ are obtained for the longitudinal and two transverse phonons, respectively. For completeness, the singlet-triplet base rate due to deformation phonons is given by
\beq
\!\!\!\Gamma^{\text{DP,ST}}_{\lambda}\approx \eta^{\text{DP}}_{\lambda}\frac{\omega^3_{Z}}{\pi^2\hbar\rho_cv^5_{\lambda}}[1+n(\omega_{Z})]\,.
\eeq
where $\{\eta^{\text{DP}}_{l},\eta^{\text{DP}}_{t1}, \eta^{\text{DP}}_{t2}\}\!=\!\{\frac{(2D_a^2+D_b^2)}{4},\frac{D_b^2}{140},0\}$ with $D_a\!=\!1.16$ eV and $D_b\!=\!-2.0$ eV for GaAs \cite{Climente_NJP13}.

When the SO mixture is large enough, corrections from single-spin relaxation channels should be included. The Rashba interaction $\HH^{(j)}_{\text{R}}\propto p_{j}^3$ couples the ground orbital to the third excited orbital and so on, whereas the Dresselhaus interaction couples the ground orbital to all the others directly or indirectly. For example, the effective left spin states $\ket{\Uparrow}$ and $\ket{\Downarrow}$ due to the Dresselhaus interaction are given by $\ket{\Uparrow}\!=\!\sum_{\sigma} a_{\sigma,0}\ket{\sigma}\ket{0,0}+\sum_{n', \sigma}a_{\sigma,n'}\ket{\sigma}\ket{n'_x,n'_y}$ and $\ket{\Downarrow}\!=\!\sum_{\sigma} b_{\sigma,0}\ket{\sigma}\ket{0,0}+\sum_{n', \sigma}b_{\sigma,n'}\ket{\sigma}\ket{n'_x,n'_y}$, where $n'=\{n'_x,n'_y\}\neq \{0,0\}$ includes all the excited orbitals, and $a_{\sigma,n}$ and $b_{\sigma,n}$ are the coefficients of the $\ket{\Uparrow}$ and $\ket{\Downarrow}$, with the corresponding spin $\sigma$ and orbital indices $n$. In this case, the piezoelectric-phonon relaxation rate is given by $\Gamma^{\Uparrow\Downarrow}\!=\!\sum_{n'}(|a^{*}_{\ua,0}b_{\da,n'}|^2+|b_{\da,0}a^{*}_{\ua,n'}|^2)\Gamma^{\text{PE}}_{\lambda}$, where the base rate
\beq
	\Gamma^{\text{PE},\Uparrow\Downarrow}_{\lambda}\approx \gamma^{\text{PE}}_{\lambda}\frac{(eh_{14})^2R^2_0}{\hbar\rho_c\epsilon^2_{r}v^5_{\lambda}}\omega^3_{Z} [1+n(\omega_{Z})]\,.\label{eq:rel_single}
\eeq
with $\{\gamma^{\text{PE}}_{l},\gamma^{\text{PE}}_{t1}, \gamma^{\text{PE}}_{t2}\}\!=\!\{\frac{4}{35},\frac{8}{105},\frac{32}{1155}\}$ and $\omega_{Z}=E_{Z}/\hbar$. 

\begin{figure}[t]
	\includegraphics[width=1.0\linewidth]{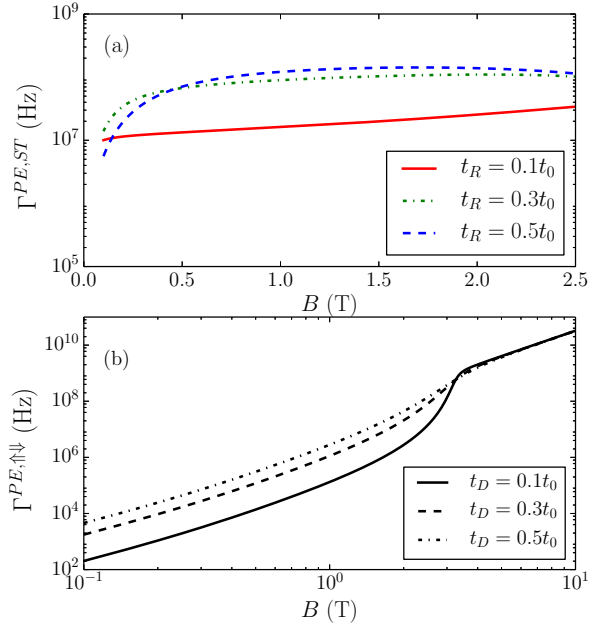}
\caption{(a) The Rashba-assisted singlet-triplet relaxation rate due to piezoelectric phonons as a function of $B$. (b) The Dresselhaus-assisted single-spin relaxation rate as a function of $B$, when the first three excited orbitals are included. $t_0=200$ $\mu$eV. The 2D confinement given by $R_0=30$ nm is about $190$ $\mu$eV. The curves in (b) meet when $B \gtrsim 3$ T as $|E_{Z}|\rightarrow t_0$ (i.e., the singlet-triplet crossing)}\label{fig:rel}
\end{figure}

In Fig.~\ref{fig:rel}, we plot $\Gamma^{\text{PE},ST}$ and $\Gamma^{\text{PE},\Uparrow\Downarrow}$ to show the field dependence in the relaxation rate. For the $\Gamma^{\text{PE},\Uparrow\Downarrow}$ plot, we include only the first three excited orbitals and the actual rate may be more enhanced. The base rate due to deformation is given by
\beq
\!\!\!\Gamma^{\text{DP},\Uparrow\Downarrow}_{\lambda}\approx \gamma^{\text{DP}}_{\lambda}\frac{R^2_0\omega^5_{Z}}{\pi^2\hbar\rho_cv^7_{\lambda}}[1+n(\omega_{Z})]\,.
\eeq
where $\{\gamma^{\text{DP}}_{l},\gamma^{\text{DP}}_{t1}, \gamma^{\text{DP}}_{t2}\}\!=\!\{\frac{(7D_a^2+D_b^2)}{84},\frac{D_b^2}{105},0\}$.

\bibliography{dot_refs_arxiv}
\end{document}